\documentclass[doublecol]{epl2}

\title{Mixed-lag synchronization in coupled counter-rotating oscillators}
\shorttitle{Mixed-lag synchronization } 

\author{Sourav K. Bhowmick$^{1,3}$\footnote{e-mail: souravb1980@gmail.com}, Dibakar Ghosh$^2$ \footnote{e-mail: dibakar@isical.ac.in}, Syamal K. Dana$^3$\footnote{e-mail: syamaldana@gmail.com}}
\shortauthor{S. K. Bhowmick, D. Ghosh and S. K. Dana}

\institute{$^1$Department of Electronics, Asutosh College, Kolkata 700 026, India.\\
$^2$Physics and Applied Mathematics Unit, Indian Statistical Institute, Kolkata 100 108, India.\\
$^3$CSIR-Indian Institute of Chemical Biology, Kolkata 700 032, India.}

\abstract{ We report mixed lag synchronization in coupled counter-rotating oscillators. The trajectories of counter-rotating oscillators has  
opposite directions of rotation in uncoupled state. Under diffusive coupling via a scalar variable, a mixed lag synchronization emerges when a parameter mismatch 
is induced in two counter-rotating oscillators. In the state of mixed lag synchronization, one pair of state variables achieve synchronization shifted in time while 
another pair of state variables are in antisynchronization, however, they are too shifted by the same time. Numerical example of the paradigmatic R{\"o}ssler oscillator 
is presented and supported by electronic experiment. }

\begin{document}
\maketitle

 The natural presence of a parameter mismatch in chaotic oscillators under diffusive coupling induces a lag synchronization (LS) \cite {rosenblum1, yclai, corron} 
instead of a complete synchronization (CS) \cite{picora}. Interestingly, the LS appears in a weaker coupling limit lower than that of CS \cite {chlai, banerjee}. 
In such a LS state, all the pairs of state variables of the coupled systems maintain a strong amplitude correlation but shifted by a common time. 
Antilag synchronization (ALS) or inverse lag synchronization \cite {guo} is another LS scenario in mismatch oscillators when two state variables are in 
antisynchronization state but with a common time shift. 
The amplitude correlation starts decreasing with decreasing coupling strength yet a phase synchronization (PS) \cite{pikovsky} emerges at a lower critical
 coupling when the phases of the coupled oscillators maintain a constant difference. 
\par We focus on LS scenario that demands more attention since it is more realistic in the larger coupling regime and has potential applications \cite {corron, bhowmick}.
 However, we consider counter-rotating oscillators as our target candidate of LS studies instead of the conventional co-rotating oscillators. In contrast to the usual LS in 
mismatched co-rotating oscillators, a mixed LS and ALS emerges in two counter-rotating oscillators under instantaneous diffusive 
coupling above a critical strength.
One pair of state variables show LS while another pair is in ALS.
Counter-rotation in oscillators is a physical reality as shown in an electronic experiment \cite {bhowmick} when two 
oscillators are characteristicially same \cite {prasad} but their trajectories rotate in opposite directions, clockwise or aniclockwise, in phase space. 
The sense of a direction in the rotation of a trajectory of a dynamical system was first introduced by Tabor \cite {tabor} which was elaborated 
further by Prasad \cite {prasad}. A general rule how to change the direction of rotation in a dynamical system was proposed by the authors \cite {bhowmick}. 
It was reported that, in two identcal counter-rotating oscillators, different pairs of state varaibles emerge into a mixed CS and antisynchronization(AS) state 
when a scalar coupling is  introduced \cite {bhowmick}.  
By introducing a mismatch in this coupled counter-rotating oscillators, the mixed lag scenario (MLS) is expected when LS and ALS coexist. 
We report, in this Letter, the details of the emergence of this MLS in mismatched counter-rotating oscillators. We use the paradigmatic 
R{\"o}ssler model \cite{rossler} for numerical demonstration and an electronic circuit for experimental evidence. 
Such a MLS can be engineered \cite {bhowmick2} in co-rotating chaotic systems by a design of delay coupling, however, to our best knowledge, 
it is not reported so far in systems under instantenous diffusive coupling where it is an emergent behavior.

In the R\"{o}ssler oscillator, the trajectory has a {\it x-y} plane of rotation while it tries to escape intermittently along the {\it z}-axis but returns back 
to the plane of rotation and it repeats. The R\"{o}ssler system may be described by,
 \begin{eqnarray} \label{rossler1} \dot{x}=Ax+f(x)+C, 
&&x \in \Re^3 \end{eqnarray} 
where $A$ is a $3\times 3$ constant matrix and represents the linear part of the system, $f : \Re^3 \rightarrow  \Re^3$ contains the nonlinear part of the system, and $C$ is a $3\times  1$ constant
matrix.

By giving a spatial rotation of the Euler angle of the {\it A} matrix, a change of direction of the trajecrory of a dynamical system is incorporated \cite{bhowmick}. 
Two counter-rotating R{\"o}ssler oscillators are thereby obtained \cite{bhowmick} and described with a scalar coupling,

\begin{eqnarray}
\label{rossler}
&&\dot{x}_i= -\omega_iy_i-z_i+\epsilon(x_j-x_i)\nonumber\\
&&\dot{y}_i= \omega_ix_i+ay_i\nonumber\\
&&\dot{z}_i=b+z_i(x_i-c)
\end{eqnarray}
\begin{figure}[ht]
\centerline{
\includegraphics[scale=0.5]{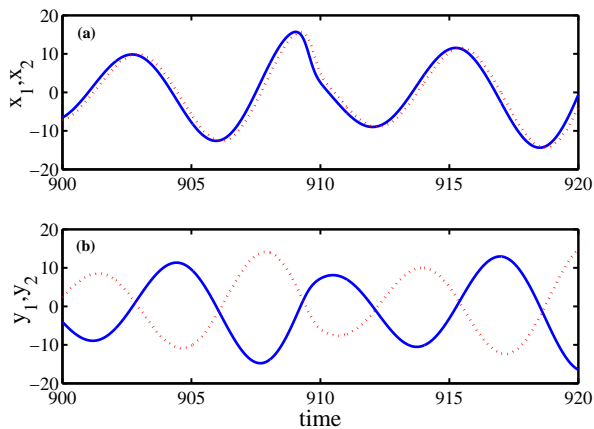}}
\caption{Time series for LS between state variable $x_1$ and $x_2$ and ALS between state variables $y_1$ and $y_2$ with $\epsilon =0.2$. 
Other parameters are unchanged.}  
\end{figure}

\begin{figure}[ht]
\centerline{
\includegraphics[scale=0.5]{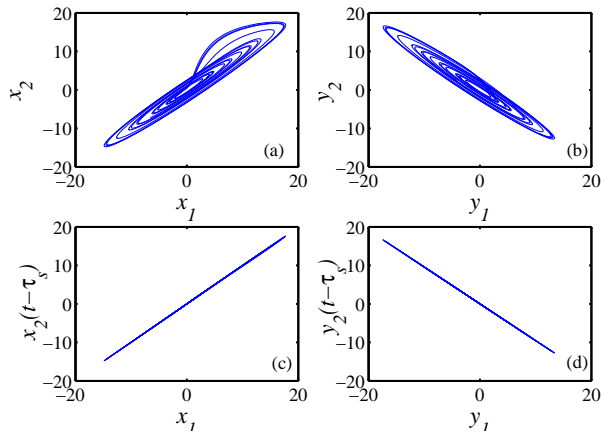}}
\caption{Synchronization manifold of the coupled system  on the plane $x_2$ vs. $x_1$ and $y_2$ vs. $y_1$ 
and delayed version plot $x_2$ vs. $x_1(t-\tau_s)$ and $y_2$ vs. $y_1(t-\tau_s)$ for  coupling $\epsilon =0.2$ 
a regime with lag synchronization with $\tau_s=0.21$.} 
\end{figure}

where $i,j=1, 2$, $i\not=j$ represent two oscillators, $\omega_1=\omega+\Delta \omega$, $\omega_2=-(\omega-\Delta \omega)$ 
and $\epsilon$ is the coupling strength and, $\omega=0.97,\Delta \omega= 0.02, a=0.165, b=0.2$ and $c=10$.
A change in the sign of $\omega_{1,2}$ parameters effectively introduce the necessary change in the Euler angle
to create counter-rotations in the system. One might not confuse here with a negative frequency since $\omega$ parameter is the representative 
frequency of the R\"{o}ssler system. It may be a different parameter not related to frequency for another system which is explained, in detail, in ref.\cite{bhowmick}.
 \par Two identical counter-rotating oscillators ($\Delta\omega=0$) emerge into a mixed synchronization (MS) state \cite {bhowmick} where the directly
 coupled variables of the plane of rotation emerge into a CS state and other variables into an AS state. For the R{\"o}ssler oscillator, 
the plane of rotation is $x-y$ and when two identical counter-rotating R\"{o}ssler oscillators are coupled via the $x$ variables, the ($x_1, x_2$)-pair of variables emerge 
into CS and the $(y_1, y_2)$-pair of variables into AS. Now for an induced mismatch in $\omega$, a MLS state emerges when a LS is found in the $(x_1,x_2$)-pair 
of variables and ALS in the ($y_1,y_2$)-pair of variables as shown in Fig. 1. The time series of $(x_1, x_2)$- and $(y_1, y_2)$- pair of variables are shown in 
Fig. 1(a) and 1(b) respectively. The $x_1(t)$ and $y_1(t)$ are plotted in solid blue lines and $x_2(t)$ and $y_2(t)$ in dotted red lines. 
In Fig.  2, $x_1(t)$ vs. $x_2(t)$ and $y_1(t)$ vs. $y_2(t)$ plots are shown which reveals LS and ALS respectively. This is confirmed when similar plots 
are made with an appropriate time shift $\tau_s$ in the $x_2(t)$ and the $y_2(t)$ variables in Fig. 2(c) and 2(d) respectively. 
This time shift is estimated by using a cross-correlation measure.
  



A cross-correlation measure $\rho$ \cite{muleti} is used here to identify the largest correlation between 
any pair of time series, say $x_1(t)$ and $x_2(t)$, for varying time shift $\tau$,

\begin{equation}
\label{2}
\rho_{x_1x_2}=\frac{<[x_1(t)x_2(t-\tau)]>}{[<x_1(t)^2><x_2(t)^2>]^{\frac{1}{2}}}
\end{equation}
where $<.>$ means a time average. The cross-correlation measure shows a largest maximum at a critical $\tau=\tau_s$ value.

As mentioned in ref. \cite{rosenblum}, the lag time for all the pairs of variables is identical in a LS state; they have a common time shift
which is also found true for the counter-rotating oscillators. For the estimation of cross-correlation between $y_1$ and $y_2$ which shows ALS,
an inverse of one of the $y$ variables is considered for the R{\"o}ssler oscillator.
The cross-correlation plot between $x_1$ and $x_2$ is shown in Fig. 3 for a coupling strength $\epsilon=0.2$.
At $\tau=\tau_s=0.21$, the correlation plot shows a largest maximum that estimates the time shift between the related state variables.
An identical time shift is found between $y_1$ and $y_2$. For the $z$ variables, the time shift is also identical but not shown here.

\begin{figure}[ht]
\centerline{
\includegraphics[scale=0.5]{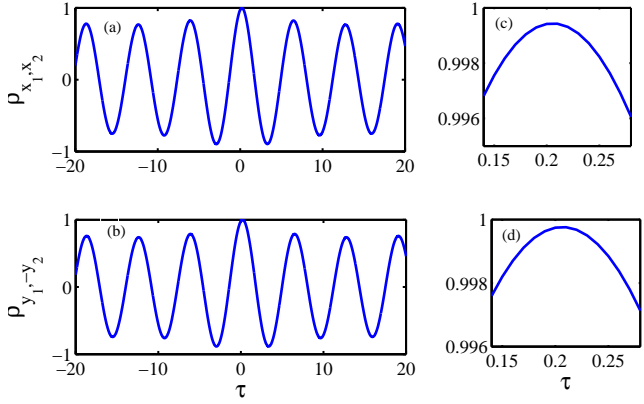}}
\caption{Cross-corelation between (a) $(x_1, x_2)$ and (b) $(y_1, -y_2)$. A closer view of the largest maximum of $\rho$
for pairs of variables (c) $(x_1,x_2)$, (d) ($y_1,y_2$). For both the cases $\tau_s$=0.21.}
\end{figure}


\begin{figure}[ht]
\centerline{
\includegraphics[scale=0.6]{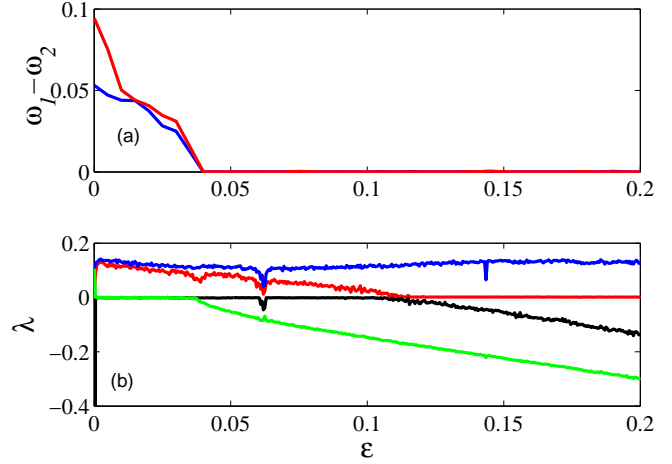}}
\caption{(a) Average frequency difference of ($x_1,x_2$) and ($y_1,y_2$) in blue lines and red lines respectively.
(b) Four lyapunov exponents are plotted with coupling strength. Average frequency difference becomes zero (a) at $\epsilon=0.04$ when 
one zero lyapunov exponent (green line) also becomes negative (b). At a larger coupling $\epsilon=0.12$, one positive exponent (red line) becomes zero and
the second zero exponent (black line) moves to a negative value.} 
\end{figure}

The average frequency difference as well as the lyapunov exponents of the coupled counter-rotating oscillators is estimated with varying coupling strength 
for identification of the critical coupling for a stable MLS. A mixed phase synchronization (MPS) is first identified for a critical coupling $\epsilon=0.04$ 
as shown in Fig. 4 when the average frequency difference goes to zero. The average frequency of an oscillator is estimated from a scalar variable  
or a time series \cite {blasius, dana} for a long run. In Fig. 4(a), we plot the average frequency differences of the $(x_1, x_2)$-variables (in blue lines) 
which are in inphase  and, the  $(y_1, y_2)$-variables (in red lines) those are in antiphase. However, the MLS is not recognizble  from the average frequency 
differene plot. 
For this, the first four lyapunov exponents of the coupled counter-rotating oscillator are plotted in Fig. 4(b) which shows two positive values and two zeros 
for small coupling. At $\epsilon=0.04$, one zero exponent moves to a negative value indicating the onset of MPS in a similar manner the PS emerges in 
co-rotating oscillators \cite {rosenblum}. At a larger coupling $\epsilon=0.12$, one positive lyapunov exponent becomes zero and the second zero exponent 
becomes negative which usually indicates an onset of LS \cite{rosenblum1, pikovsky1, osipov} and, in our case, it is an emergence of a MLS state. 
\par Now we present an experimental verification of this MLS scenario using an electronic analog of the piecewise R{\"o}ssler \cite{carrol} model,

\begin{eqnarray}
\label{piecewise}
&&\dot{x}= -\alpha x-\omega \beta y-\lambda z\nonumber\\
&&\dot{y}=  \omega x+\gamma y-0.02z\nonumber\\
&&\dot{z}=g(x)-z
\end{eqnarray}
where
\begin{center}
$g{(x)}= 0$  if $x\leq 3$\nonumber\\  
  $\;\;\;\;\;\;\;\;\;\;\;\;\;\;\;\;\;\;\; =\mu (x-3)$  if $ x\geq 3$
\end{center}
We construct two counter-rotating circuits of the piecewise linear R{\"o}ssler model as detailed in ref. \cite{bhowmick}. For counter rotation, 
we cosider $\omega =\pm 1$ and other parameters are $\alpha=0.05,  \beta =0.5, \lambda =1.0,\mu =7, \gamma =0.204$.  
The mismatch is given in parameter $\beta $. For two oscillators, the value of $\beta $ are $0.5$ and $0.45$.

\begin{figure}[ht]
\centerline{
\includegraphics[scale=0.25]{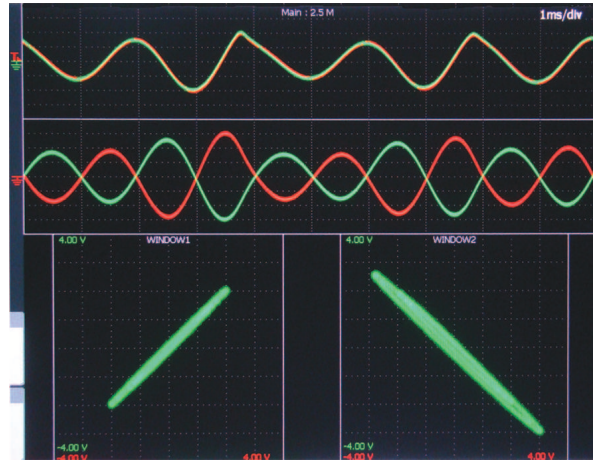}}
\caption{(Color online) Oscilloscope pictures: lag synchronization between $x_1(t)$ and $x_2(t)$ in upper panel, antilag synchronization between
$y_1(t), y_2(t)$ in middle panel. Lag synchronization is confirmed in $x_1(t) vs. x_2(t)$ plot (left lower panel) and antilag in $y_1(t) vs. y_2(t)$ 
plot (right lower panel).} 
\end{figure}

\begin{figure}[ht]
\centerline{
\includegraphics[scale=0.5]{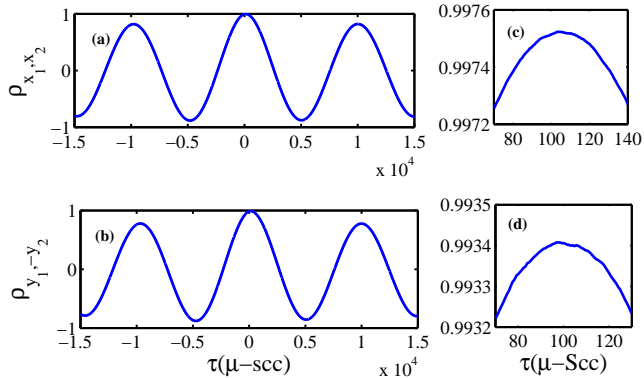}}
\caption{Experimental cross-corelation plot for (a)($x_1, x_2$)-pair, (b) ($y_1,-y_2$)-pair of variables.
A closer view of the largest maxima both cases are shown in (c) and (d) respectively. 
For both the pair of variables, the time shift is approximately identical, $\tau=100 \mu$s. within the experimental error limit.}
\end{figure}

Figure 5 (oscilloscope pictures) shows experimental evidence of the MLS. The oscilloscope pictures (Yokogawa DL 9140, 
4-channel, 1GHz, 5 GS/s) show pairs of time series, ($x_1(t), x_2(t)$) and ($y_1(t), y_2(t)$) in upper panel and middle panel respectively
 which are in in-phase and antiphase state but with a time shift.  The phase plane plot of ($x_1(t) vs. x_2(t)$) and ($y_1(t) vs. y_2(t)$) are shown 
in the lower panels of Fig. 5. 
The lag time of experimental time series is estimated using the cross-correlation measure. The cross-correlation estimates for both the pair of time 
series ($x_1,x_2$) and $(y_1,y_2$) are presented in Figs. 6(a) and 6(b) respectively. A closer view of cross-correlation for both the cases are
plotted in Fig. 6(c) and 6(d) and confirms a largest maxima at a critical lag time $100\mu$s.

In summary, we reported a mixed-lag scenario in mismatched counter-rotating oscillators under diffusive scalar coupling. In this mixed-lag synchronizaton regime,
one pair of state varaibles of the coupled system emerged into lag synchronization while another pair established anti-lag synchronization. 
This is a unique and a novel feature of chaotic systems, not reported so far, to our best knowledge. We used average frequency difference
 and lyapunov exponent estimation to identify the critical coupling strength for the onset of the stable mixed-lag synchronization. We used a cross-correlation 
measure of the related state variables to estimate the amount of time shift or lag between the different pairs of state variables of the coupled system. 
We provided both numerical and experimental demonstration of this mixed lag scenario using a numerical example of the R\"{o}ssler system and electronic analog of
counter-rotating oscillators of a piecewise linear R\"{o}ssler system.
 
\section{Acknowledgements}
S.K.B. is supported by the BRNS/DAE, India (2009/34/26/BRNS). S.K.D. acknowledges support by the CSIR Emeritus scientist scheme.

\bibliography{aipsamp}
\end{document}